\documentclass[twocolumn,epsfig,prl,reprint, showpacs]{revtex4}

 \pdfoutput=1
 
\usepackage{amsmath}
\usepackage{amsfonts}
\usepackage{amssymb}
\usepackage{graphicx}
\usepackage{color}

\usepackage{hyperref}
\hypersetup{colorlinks,linkcolor=blue,filecolor=green,urlcolor=blue,citecolor=blue}

\newcommand{\be}{\begin{equation}}
\newcommand{\ee}{\end{equation}}

\newcommand{\sysb}{\left\{\begin{array}}
\newcommand{\syse}{\end{array}\right.}

\newcommand{\ag}[1]{}
\newcommand{\jm}[1]{\textcolor{blue}{#1}}

\begin{document}

\title{Casimir forces and quantum friction from Ginzburg radiation in atomic BECs}

\author{Jamir Marino,${}^{1,2}$ Alessio Recati,${}^{3, 4}$ and Iacopo Carusotto${}^{4}$}

\affiliation{$^1$ Institute of Theoretical Physics, TU Dresden, D-01062 Dresden, Germany}

\affiliation{$^2$ Institute of Theoretical Physics, University of Cologne, D-50937 Cologne, Germany}

\affiliation{$^3$ Physik Department, TU M\"unchen, James-Franck-Stra{\ss}e 1, 85748 Garching, Germany}

\affiliation{$^4$ INO-CNR BEC Center and Dipartimento di Fisica, Universit\`a di Trento, 38123 Povo, Italy.}

\begin{abstract}
We theoretically propose an experimentally viable scheme to use an impurity atom in an atomic Bose-Einstein condensate, in order to realize condensed-matter analogs of  quantum vacuum effects.
In a suitable atomic level configuration, the collisional interaction between the impurity atom and the density fluctuations in the condensate can be tailored to closely reproduce the electric-dipole coupling of quantum electrodynamics. By virtue of this analogy, we recover and extend the paradigm of electromagnetic vacuum forces to the domain of cold atoms, showing in particular the emergence, at supersonic atomic speeds, of a novel power-law scaling of the Casimir force felt by the atomic impurity, as well as the occurrence of a quantum frictional force, accompanied by the Ginzburg emission of Bogoliubov quanta. Observable consequences of these quantum vacuum effects in realistic spectroscopic experiments are discussed.
\end{abstract}



\date{\today}

\maketitle

\emph{Introduction --} 
One of the most exciting features of modern quantum field theory consists of the non trivial structure of the vacuum state and of the associated zero-point fluctuations. Among their most celebrated observable consequences in the electromagnetic (EM) context, we can mention the static Casimir force between neutral objects~\cite{Casimir} experimentally demonstrated in~\cite{Suk93, Dru03, Ob07}, and the anticipated dynamical Casimir emission of correlated pairs of photons by suitably accelerated neutral objects~\cite{DCE, Dod2010}.

The physics underlying such pair emission processes is perhaps cleanest in the case of Ginzburg radiation from neutral but polarizable objects uniformly moving at superluminal speeds~\cite{Ginz}. The so-called anomalous Doppler effect allows in fact an atom moving in a medium at $v>c/n$ ($n$ being the medium refractive index) to jump from its ground state to an excited state by emitting (and not absorbing) a photon.

So far, the challenge of making a mirror or a neutral particle to move at ultrarelativistic speeds in a medium has prevented a direct experimental observation of these fascinating emission processes by physically moving massive objects or atoms. As a result, the only available experimental observations of dynamical Casimir emission \cite{DCE, DCE2} were based on EM analogs of the moving mirror using a tunable reflecting element in a superconducting device ~\cite{Wilson2011, Paraoanu2013}. 

In this Letter we develop and exploit a general framework to  study a wide variety of quantum vacuum effects 
using state-of-the-art cold atom technology. Following the same spirit of the quest for analog Hawking radiation in analog 
models of gravity~\cite{Living} and building on recent works on analog Casimir forces~\cite{Recati2005a, Recati2005b, Fleish2005, Roberts2005, Henkel2008, Kamenev2014} and analog Unruh temperature~\cite{Reznik} in quantum 
fluids, we propose to employ a dilute Bose-Einstein condensate (BEC) as the medium and Bogoliubov sound waves in place of EM~waves as the quantum field. The requirement for an ultrarelativistic motion is in this way replaced by a much more accessible condition involving the speed of sound in the condensate, typically in the \emph{cm/s} range~\cite{LPSS}. 
In contrast to previous works on analog Casimir forces in quantum fluids where the coupling is a charge-like one~\cite{Recati2005a, Recati2005b, Fleish2005, Roberts2005, Henkel2008, Kamenev2014}, here the impurity behaves as a neutral two-level atom coupled to the EM field, i.e.  with a vanishing charge but a non-vanishing polarizability.

\begin{figure}[t!]
\includegraphics[width=6.0cm]
{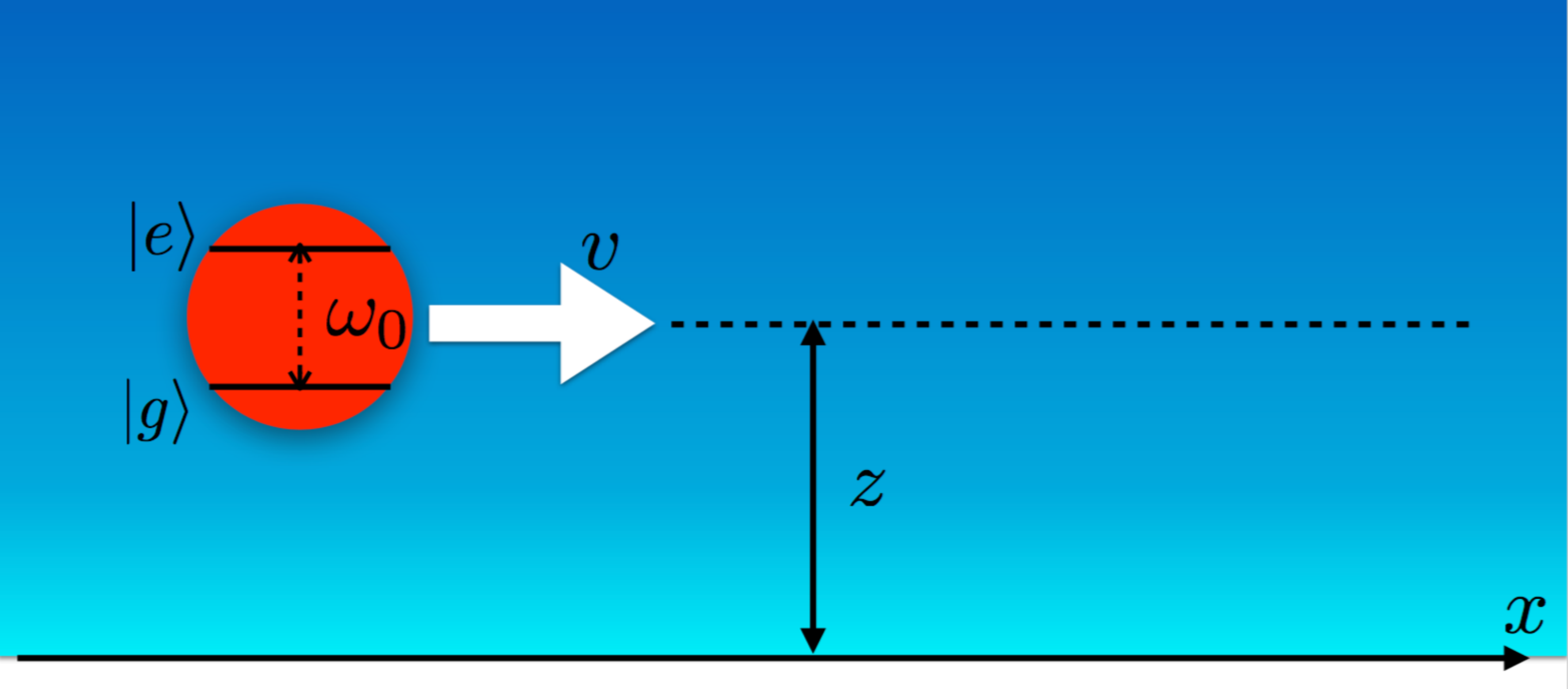}
 \caption{(Color online) Schematic representation of the moving impurity with internal frequency $\omega_0$, and the plate on the $z=0$ plane, confining the condensate (in blue).} 
\label{Fig0}
\end{figure}

As a first and most promising application of our general theory, we will show how simple atomic spectroscopy tools may provide crucial information on the Ginzburg emission from a uniformly moving impurity. As soon as its speed exceeds the speed of sound, the atomic ground state acquires a detectably finite linewidth due to spontaneous excitation processes to the excited state and experience a mechanical friction force in response to the Ginzburg emission. When the {supersonic} moving impurity is located in the vicinity of the edge of the BEC, it experiences an analog of the  {zero-temperature} Casimir force with  a novel power law scaling, and an analog of the Casimir friction of quantum electro-dynamics (QED)~\cite{Scheel09, Vol2007, Pendry2010, Barton2010, Intravaia14, Intravaia15, Intravaia16}.
{This novel  phenomenology  sets apart our results from previous studies  of Casimir forces in condensed matter settings~\cite{Fleish2005, Recati2005a, Recati2005b, Roberts2005, Henkel2008, Kamenev2014}}.
Experimentally realistic protocols to highlight these effects as frequency shifts and broadenings of the atomic transition lines are put forward, as well as estimates for the observability of our predictions.

\emph{The BEC platform for quantum vacuum effects --} 
Inspired by the atomic quantum dot idea originally introduced in~\cite{RecatiQD}, we consider an impurity consisting of a two-level ($1$ and $2$) atom, immersed in a three-dimensional atomic Bose-Einstein condensate at very low temperature. The motion of the impurity is supposed to be externally imposed by a tightly confining and uniformly moving trap potential, so that its only degrees of freedom are the internal ones. The impurity is assumed to be illuminated by a monochromatic external electromagnetic field at the frequency $\omega_L$ close to resonance with the $1\to 2$ transition, $\omega_L\simeq\omega_{21}$, with a (real and positive) Rabi frequency $\omega_0$. Employing the rotating wave approximation (RWA) for the atom-field interaction, we can write the system Hamiltonian in the form 
\begin{eqnarray}
H{(t)}&=& H_{C}+H_A{(t)}=\int\!d^3\mathbf{r}\left(\hbar^2 \frac{\nabla\hat{\psi}^\dagger\,\nabla\hat{\psi}}{2m}+\lambda\hat{\psi}^\dagger\hat{\psi}^\dagger\hat{\psi}\hat{\psi} \right)\nonumber \\
&+& \hbar\omega_{21}|2\rangle\langle 2|-\left[\frac{\hbar\omega_0}{2}\,e^{-i\omega_L t}\,|2\rangle\langle 1|+\textrm{h.c.} 
\right] \nonumber \\
&+&\sum_s g_s \hat{\rho}(\mathbf{r}_A(t))|s\rangle\langle s|,
\label{eq:H_C+A}
\end{eqnarray}
{where the atomic Bose gas density reads $\hat{\rho}(\mathbf{r})=\hat{\psi}^\dagger(\mathbf{r})\hat{\psi}(\mathbf{r})$ in terms of the atomic Bose field $\hat{\psi}(\mathbf{r})$ and needs evaluating at the (generally) time-dependent impurity position $\mathbf{r}_A(t)=\mathbf{v}t$.}
All atom-atom interactions are taken to be local in space; $\lambda$ is the interaction constant between condensate atoms, while $g_{s}$ are the ones between the impurity in the $s=1,2$ state and the condensate atoms {of mass $m$}. Processes where the impurity atom switches state under the effect of collisions with the condensate are forbidden by angular momentum conservation.
The impurity Hamiltonian written in the second and third lines of (\ref{eq:H_C+A}) can be made time-independent by {switching to the interaction picture the light-matter interaction Hamiltonian  and choosing a reference frame comoving with the uniformly moving impurity at speed $\mathbf{v}$}.

Using the rotated $|g,e\rangle=\frac{1}{\sqrt{2}}(|1\rangle\pm |2\rangle)$ basis and defining $g_\pm=\frac{1}{2}(g_1\pm g_2)$, 
and the detuning $\delta=\omega_L-\omega_{21}$, we obtain
\begin{equation}
 H_A=-\frac{\hbar \omega_0}{2} \sigma_z + \frac{\delta}{2}\,\sigma_x + \hat{\rho}(\mathbf{r}_A)[g_+ + g_- \sigma_x],
 \label{eq:H_A}
\end{equation}
where $\sigma_{x,z}$ are the usual Pauli matrices. In this rotated basis, the Rabi frequency $\omega_0$ determines the 
splitting between the $|g,e\rangle$ states, while the detuning $\delta$ gives a coupling term. The square bracket describes 
the interaction with the condensate: the first term proportional to $g_+$ resembles the coupling of a static charge to a field,
while the second term proportional to $g_-$ resembles a standard electric dipole coupling~\cite{CCT4}. In order to isolate the latter coupling, one has to impose the condition $g_+=0$ for analog charge neutrality, which can be satisfied with suitably chosen internal atomic states and a fine-tuning of the $g_{1,2}$, e.g., via Feshbach resonance. 
For instance,  two hyperfine levels of potassium $^{40}$K interacting with rubidium atoms $^{87}$Rb (see for instance Refs. \cite{firenze}), realise, for  magnetic fields of around 700 Gauss, the condition $g_+=g_{1}+g_2=0$, while showing a relatively large dipole coupling $g_-$ corresponding to a scattering length of a few $10^2$  Bohr radii. 
The analogy with quantum electrodynamics becomes clear if we split the atomic density $\hat{\rho}(\mathbf{r})$ into its 
(spatially homogeneous) average value $\rho_0$ and small fluctuations $\delta \rho(\mathbf{r})$, and we assume that the 
detuning $\delta$ exactly compensates in Eq.~\eqref{eq:H_A} the coupling to the average density, $\delta/2+g_-\rho_0=0$.
Under this assumption, the coupling of the impurity to the condensate has analogous form, 
$g_-\,\sigma_x\,\delta \rho(\mathbf{{r_A}})$, of a two-level atom dipole-coupled to the quantum EM field {at its position $\mathbf{r}_A$}.

Within the Bogoliubov theory of small excitations on top of the (dilute) condensate~\cite{LPSS}, 
density fluctuations can in fact be written in a form $\delta \rho(\mathbf{r},t)=\sqrt{\rho_0}\int \frac{d^3k}{(2\pi)^3} (u_{k}+v_{k})( e^{i\mathbf{k}\cdot\mathbf{r}}b_{\mathbf{k}}(t)+e^{-i\mathbf{k}\cdot\mathbf{r}}b_{\mathbf{k}}^{\dagger}(t))$,
that closely resembles the quantized electric field of QED in terms of bosonic operators 
$b_{\mathbf{k}}$ ($b^\dagger_{\mathbf{k}}$) satisfying the usual Bose commutation rules 
$[b_{\mathbf{k}},b^\dagger_{\mathbf{k'}}]=(2\pi)^3 \delta^{(3)}(\mathbf{k}-\mathbf{k}')$, which destroy (create) a 
collective excitation of the condensate with wavevector $\mathbf{k}$. 
$\delta \rho(\mathbf{r},t)$ is written in Heisenberg representation and the time evolution of the operators $b_{\mathbf{k}}(t)\equiv b_{\mathbf{k}}(0)e^{-i\hbar\omega_k}$, is dictated by the Bogoliubov dispersion relation. {In the laboratory frame where the condensate is at rest, this reads}
$\hbar\omega_k=c_s k\sqrt{1+(k\xi)^2}$, where both the speed of sound $c_s=\sqrt{\mu/m}$ and the healing length 
$\xi=\frac{\hbar}{2\sqrt{m\mu}}$ are given in terms of the chemical potential $\mu=\lambda\rho_0$ of the condensate~\cite{LPSS}. The spectrum is linear only at small momenta $k\ll 1/\xi$, while for high wavevectors the dispersion deviates upwards and 
tends to the non-relativistic massive particle $E_k={\hbar^2k^2}/{2m}$ one. {As usual in a frame comoving with the impurity at speed $\mathbf{v}$, the Bogolyubov dispersion gets Doppler shifted to $\hbar\omega_{k}-\mathbf{k}\cdot\mathbf{v}$}.
Notice that in $\delta \rho(\mathbf{r},t)$ we have $(u_k+v_k)^2=E_k/\hbar\omega_k$, {since the Bogoliubov coefficients read~\cite{LPSS} $u_k^2=1/2(\zeta_k/(\hbar\omega_k)+1)$, $v_k^2=1/2(\zeta_k/(\hbar\omega_k)-1)$ with $\zeta_k=E_k+\mu$. }



\emph{Casimir forces for a static impurity --}  As a first example of non-trivial quantum fluctuation effects, 
we determine the Casimir force felt by a dressed impurity \jm{at rest ($v=0$)} in its ground state $g$ close due the condensate's
surface. For the sake of simplicity we assume a flat surface located along the $z=0$ plane, see Fig. \ref{Fig0},
imposing the condition $\delta \rho(\mathbf{r})|_{z=0}=0$. This condition closely resembles the one currently used in QED to describe a perfectly reflecting mirror~\cite{books}. 
As we are focusing on low energy phonons, we expect that it provides a reasonable approximation of more realistic boundary conditions at the condensate surface~\cite{henk}. 

The Casimir force is given by the negative derivative along $z$ of the energy shift due to the coupling of the 
impurity with the quantum fluctuations of the density. 
As customary in the literature on the Casimir effect in either scalar or complete QED~\cite{Wylie, books}, we determine the latter within second order perturbation theory {(see Supplemental Material for further details)}.
Assuming  $\Omega\equiv\hbar\omega_0/\mu\ll1$ (i.e., tuning to a small value the Rabi frequency), there is a clear scale separation between  the healing length 
$\xi$ and the length scale $l_C={c_s}/{\omega_0}={\xi}/{\Omega}$, which sets apart the near and far zone regimes of the Casimir 
effect. 
In the \emph{near zone}, $z\ll l_C$, the force (directed by symmetry along the $\hat{z}$ axis) scales as 
$F^g_z\simeq-g_-^2/(4\pi^2 \lambda l_C z^3)$, while in the \emph{far zone} regime, $z\gg l_C$, the scaling qualitatively changes into 
 $F^g_z(z)\simeq-(g_-^2l_C)/(\pi^2 \lambda z^5)$.
In contrast to the EM case, in our cold atom set-up the Rabi frequency can be tuned to zero, which gives $\omega_0=0$, and an exponentially decaying Casimir force $F^g_z\propto e^{-2z/\xi}$ (see also Refs. \cite{Recati2005a, Fleish2005}). Henceforth, the internal structure of the atomic impurity is a crucial ingredient in order to mimic the algebraic decay of the EM Casimir force at large distances. 
The similarity with EM Casimir effect extends to the analog Lamb shift experienced by an impurity at rest in the excited $e$ 
state at a distance $z$ from the condensate edge. In the far zone ${z\gg l_C}$, its $z$-dependence gives a  force 
along the $\hat{z}$ direction equal to   $F^e_z(z)=g_-^2/(16\pi \lambda l_C^4)(\sin(z/{l_C}))/(z/l_C)$,
which displays spatial oscillations on the top of an algebraic decay. Such oscillations are a signature of the stationary 
wave arising from  emission of quanta from the impurity and their subsequent absorption after reflection on the condensate 
edge.

\emph{Ginzburg radiation --} 
For an impurity moving along the $\hat{x}$ direction parallel to the planar edge at a speed  $v\lesssim c_s$, 
the above scenario remains unaltered, with identical scalings of the Casimir forces and no possibility for spontaneous 
excitation of the impurity. On the contrary, a de-excitation mechanism  analog of spontaneous emission exists for the 
excited state, and for $v=0$ it occurs with the transition rate $\hbar\Gamma^e=g_-^2/(32\pi \lambda l_C^3)$.

\begin{figure}[t!]
\includegraphics[width=8.8cm]
{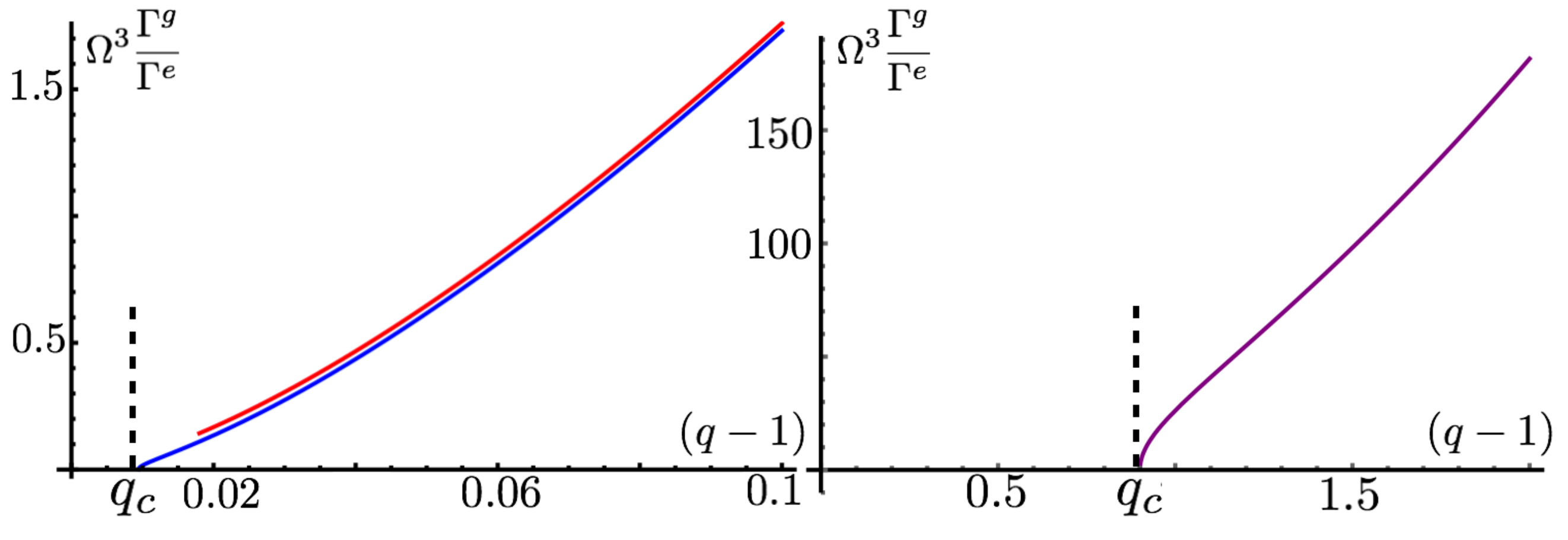}
 \caption{(Color online) Numerical evaluation of Eq. \eqref{gamma} (left panel, blue line) versus Eq. \eqref{gammaappro} (left panel, red line) in units of $\Gamma_e$, for $\Omega=10^{-3}$. The right panel shows the transition rate Eq. \eqref{gamma} for $\Omega=1$. }

\label{Fig1}
\end{figure}
The physics is dramatically different when the impurity moves at a (uniform) supersonic speed, $v\gtrsim c_s$. While the ``charge neutrality'' of the impurity rules out Bogoliubov-Cherenkov emission~\cite{CarusottoPRL2006,Astrakharchik}, the anomalous Doppler effect allows for it to jump from the ground to the excited state while emitting a Bogoliubov phonon and 
still conserving energy. This effect is the BEC analog of the excitation mechanism originally introduced by Ginzburg for 
superluminal moving particles~\cite{Ginz}. The transition rate $\Gamma^g$ is easily determined by using the Fermi 
Golden Rule. Assuming both the impurity and the Bogoliubov field initially in the ground
state {and working in the frame comoving with the impurity where the Hamiltonian is time-independent,} one has
\begin{equation}\label{gamma}
\hbar\Gamma^{g}=2\pi g_-^2\rho_0\int\frac{d^3k}{(2\pi)^3}(u_k+v_k)^{2}\delta(\hbar\omega_0+\hbar\omega_k-\hbar\mathbf{k}\cdot\mathbf{v} ),
\end{equation}
{where the $\delta$-function encoding energy conservation involves the Doppler-shifted Bogolyubov energy.}
After angular integration, the condition for a mode to give a non-vanishing contribution to the 
transition rate becomes $\hbar\omega_k-\hbar kv<-\hbar\omega_0$, 
and Eq. \eqref{gamma} yields,  for
$\Omega\ll1$ and $q\simeq q_c$, 
\be\label{gammaappro}
\hbar\Gamma^{g}\simeq\frac{g_-^2}{12\pi \lambda \xi^3}\left((2(q-1))^{3/2}-\frac{\Omega^3}{8(q-1)^3}\right).
\ee
Note that the finite frequency of the impurity transition $\omega_0>0$ is responsible for an
upward (tiny for $\Omega\ll1$) shift of the velocity threshold, $q_c$, of the Ginzburg excitation mechanism above the Landau critical velocity 
$q_{\rm Landau}=1$ proper for impurities with no internal structure and a ``charge-like'' coupling to the 
condensate~\cite{Astrakharchik}. Examples of plots of $\Gamma^g$ in different parameter regimes are shown in Fig. \ref{Fig1}.

\emph{Supersonic Casimir force --} When the condensate is bounded by a flat interface on the plane $z=0$, we can study the effect of supersonic velocities on the Casimir force. Focussing on the large distances regime $z\gg q l_C$, and supersonic speeds (\jm{\emph{ss}}) $q\gg 1$ (with $\Omega\ll1$), we obtain, for the $\hat{z}$ component of the force, the expression $F^{g,ss}_z(z)\simeq-\frac{g_-^2}{4(2 \pi)^{3/2}\lambda}\frac{1}{q^5 l_C^4}\left(\frac{q l_C}{z}\right)^{\frac{3}{2}}\sin\left(\frac{z}{q l_C}+\frac{\pi}{4}\right).
$
%
While this result maintains the characteristic oscillatory pattern (albeit with a different spatial frequency) of the Casimir force for an excited impurity, $F^e_z(z)$,
 the large distance algebraic decay of $F^{g,ss}_z$ is remarkably different, indicating that exciting a ground state impurity through the Ginzburg mechanism is not
equivalent to starting with an impurity at rest in its excited state. The different oscillatory pattern and scaling in the 
supersonic regime is a consequence of the presence of a new scale in the system, $q l_C$. 
Numerically we find that the same leading scaling persists for impurity speeds close to the critical threshold for 
activation of the Ginzburg mechanism, $q\simeq q_c$.  We notice that similar features were   found  for  Casimir interactions in relativistic accelerated backgrounds~\cite{Rizzuto, Rizzuto2, Marino}, where the new length scale is set by the acceleration.


Analogously, also the broadening $\Gamma^g$ of the impurity ground state, acquires a $z$-dependent contribution
$\gamma^g (z)$. At the leading order for $q\simeq q_c$, and $\Omega\ll1$, we find the leading scaling behaviour with $z$
\be
\hbar\gamma^g(z)\simeq-\frac{2g_-^2}{\lambda\xi^3}\sqrt{\Omega}\left(q-1\right)^{3/2}\frac{\xi}{z}\sin\Big(2\frac{z\sqrt{q-1}}{\xi}\Big),
\ee
for distances $z\gg q \Omega l_C$.\\
%
\emph{Quantum friction --} As we have seen above, the excitation process for a ground state impurity moving at supersonic 
speeds is due to the emission of Bogoliubov quasi-particles with a non-trivial angular dependence determined by the Dirac 
delta factor in Eq. \eqref{gamma}. In order to keep the motion at constant velocity, an external force must be applied to 
the impurity to compensate for the corresponding momentum change. Computing the time-averaged rate of momentum variation 
per unit of time and focussing on its component along the direction of motion $\hat{x}$, 
we find an expression for the frictional force, $\Phi_x(z)=\mathcal{F}_x^g+f_x^g(z)$, where
\be\label{power}
\begin{split}
&\mathcal{F}_x^g=\frac{\pi g_-^2\rho_0}{\hbar}\int \frac{d^3k}{(2\pi)^3}\hbar k_x(u_k+v_k)^2\delta(\hbar\omega_{0}+\hbar\omega_k-\hbar\mathbf{k}\cdot\mathbf{v}),\\
&f_x^g(z)=-\frac{\pi g_-^2\rho_0}{\hbar}\int \frac{d^3k}{(2\pi)^3}\hbar k_x(u_k+v_k)^2\times\\
&\qquad\qquad\qquad\times\delta(\hbar\omega_{0}+\hbar\omega_k-\hbar\mathbf{k}\cdot\mathbf{v})\cos(2k_zz),
\end{split}
\ee
with  $k_z$ ($k_x$) the momentum component along the  $\hat{z}$ ($\hat{x}$) direction.
The fact that the dissipative frictional force $\Phi_x(z)$ has a purely quantum nature opens the door to a realization of the analogue of a Casimir frictional force in our condensed-matter setting%
 ~({notice that  the usual  EM  frictional force  originates  between two polarizable bodies in relative motion~\cite{Vol2007}.})
~Indeed, like for the conservative Casimir forces ($F^{g}_z$ and $F^{g,ss}_z$), the finite extension of the condensate introduces a distance dependent frictional force $f_x^g(z)$, in addition to its constant component $\mathcal{F}_x^g$, which can  be analytically evaluated for $\Omega\ll1$, with the result
$\mathcal{F}_x^g\simeq\frac{g_-^2}{16\pi \lambda q^2\xi^4}\left((q^2-1)^2+\Omega\left(q\sqrt{q^2-1}-\sinh^{-1}(\sqrt{q^2-1})\right)\right)$. As a consistency check, it is easy to verify that this formula reduces to the usual expression~\cite{Astrakharchik} for the dissipative force for impurities with no internal structure 
 in the $\omega_0=0$ limit.
On the other hand, the distance-dependent component of the friction scales at large distances $z\gg q \Omega l_C$ as
 \begin{equation}\label{Cfric}
f_x^g(z)\simeq-\frac{g_-^2}{\lambda\xi^4}\sqrt{\Omega}(q-1)^2\frac{\xi}{z}\sin\Big(2\frac{z\sqrt{q-1}}{\xi}\Big),\end{equation}
where,  as for  $\gamma_g(z)$,  the computation has been done for $\Omega\ll1$ and $q\simeq q_c$. For large velocities $q\gg 1$, we find the same scaling with distance, and a 
modified  oscillatory frequency, $f_x^g(z)\propto \frac{1}{z} \sin\left(4\frac{zq}{\Omega l_C}\right)$.
The supersonic conservative, $F^{g,ss}_z$, and frictional, Eq.  \eqref{Cfric}, Casimir forces display different oscillatory frequency and leading scaling behaviour with $z$, since they  contribute to different physical mechanisms, respectively the Lamb Shift and  the broadening of the atomic spectral lines of the impurity atom. Moreover, we observe that while the Casimir friction effect vanishes for $q<q_c$, the conservative Casimir force has a different  scaling with distance above and below $q_c$.

\emph{Optical measurements --} 
So far,  experimental measurements of the (static) Casimir effect has focussed on direct measurements of the mechanical force felt by the object, which requires an extreme experimental control over all sorts of mechanical noise and systematic effects~\cite{Casimir, Bordag}. One of the key advantages of our proposal is the possibility of detecting the Casimir effect by probing transitions between atomic states and looking for frequency shifts associated to the Casimir energy instead of measuring the mechanical force.
The simplest choice to this purpose is to study the $g\to e$ transition. However, as the $e$ state is naturally dissipative by emission of phonons in the condensate, even for a subsonically moving impurity, it can be advantageous to probe another transition from the state $g$ of interest towards some long-lived state with negligible collisional interactions with the condensate. In this way, the Casimir energy for the lower state $\Delta^g(z)$
is visible as a distance-dependent shift of the very narrow resonance line with no intrinsic limit to the precision of the frequency measurement.
On the other hand, the Ginzburg effect on a super-sonically moving impurity can be observed from the increased linewidth of the transition due to the broadening $\Gamma_g(z)$ of the $g$ state. While useful to detect the Ginzburg effect, such a broadening is of course a serious hindrance against a spectroscopic detection of the Casimir energy for a supersonically moving impurity as one typically has $\Delta^g \ll \Gamma^g$. Nevertheless, since $\gamma^g(z)$ and $f^g_x(z)$ scales with the same power law at large distances, the spectroscopic measurement of the line-width gives access to the scaling of the analog Casimir frictional force. 
{For the $^{40}$K~--~$^{87}$Rb mixture discussed above, we have $g_-/\lambda\simeq6$~(see Ref.~\cite{firenze}), and, even for tiny supersonic speeds, $q\simeq 1.2$~(we set $\Omega=0.1$), we find sizeable signatures of the Ginzburg effect:  $\hbar\Gamma^g/\mu\simeq0.01$ and $\gamma^g(z)/\Gamma^g\simeq1$, in the dilute limit $n\xi^3\simeq10$ and for distances $z/\xi\simeq10$. Of course, the magnitude of these effects increases significantly for larger supersonic speeds.}

\emph{Conclusions --} In this Letter, we have proposed a cold atom set-up where an optically dressed impurity embedded in an atomic BEC and coupled to its Bogoliubov sound modes, serves as a condensed-matter analog of a neutral two-level atom dipole-coupled to the quantum EM field. We have obtained a number of results on the analog Ginzburg emission from a supersonically moving impurity and the analog {zero-temperature} Casimir forces that occur when the impurity is located next to the geometrical edge of the condensate. 
{At   temperatures $k_B T\ll\hbar\omega_0$ and $v<c_s$, we expect that, for distances $l_C\ll z \ll l_T$ ($l_T\propto 1/T$  is the de-Broglie thermal wavelength),  the far zone scaling holds
while, for $z\gg l_T$,  a scaling reminiscent of the one in near zone  ($F_z^g\sim T/z^3$) is expected, following  a phenomenology analogous to EM Casimir forces at finite temperatures \cite{books}.
At supersonic speeds,  the scenario becomes more intricate, and it  constitutes an interesting perspective direction of this work.}

Our results illustrate also the potential of our quantum fluid platform as a quantum simulator~\cite{emulator} of quantum field theories:  the tunability of the sound speed, of the impurity parameters, of the condensate dimensionality and geometry, allows to access regimes otherwise unaccessible to direct QED experiments and explore open questions concerning quantum vacuum forces in novel regimes of supersonic and/or accelerated motion~\cite{Birrell, Unruh, Cal, Marino} and/or of a driven-dissipative quantum fluid~\cite{CarusottoRMP2013}.

\emph{Acknowledgments.}  We acknowledge fruitful discussions with V. M. Agranovich, D. Dalvit, C. Henkel, F. Intravaia,  R. Passante, L. Rizzuto. JM and AR acknowledges support from the Alexander von Humboldt Foundation. This work has been supported by the ERC through the QGBE grant (IC and AR), by the EU-FET Proactive grant AQuS, Project No. 640800 (IC), and by the Autonomous Province of Trento (IC and AR), partially through the project "On silicon chip quantum optics for quantum computing and secure communications" ("SiQuro").

\onecolumngrid
\appendix
\section{Appendix}
\section{Computation of the Casimir force}
In order to extract the Casimir force on the atomic-impurity embedded in the BEC, we follow standard methods already employed to compute the atom-plate scalar EM Casimir potential (see for instance Ref. [35] of the main text).

\subsection{Scaling for an impurity at rest}
We consider the impurity at rest at a distance $z$ from  the perfectly reflecting plate in the $x\hat{O}y$ plane, and coupled with the Bogolyubov field through a dipolar-like term $V=g_-\sigma^x\delta \rho(\mathbf{r})$.
In order to develop a perturbation theory in $g_-$, we need the correlation function of the atomic impurity in its ground state, $|g\rangle$,
\begin{equation}\label{atomcorr}
\langle g|\sigma_x(t)\sigma_x(t')|g\rangle=e^{i\omega_0(t'-t)},
\end{equation}
and the $z$-dependent part of the two-point correlation function of $\delta\rho(\mathbf{r},t)$ -- evaluated on the vacuum state of the Bogolyubov field and on the  space-time location of the impurity,
\begin{equation}\label{correlatore}
\begin{split}
\langle \delta\rho(\mathbf{r},t)\delta\rho(\mathbf{r}',t')\rangle&=
-\rho_0\int_0^{2\pi} d\varphi\int_0^{\pi} d\theta\sin\theta\int_0^\infty \frac{dk}{(2\pi)^3} k^2 (u_k+v_k)^2e^{i\hbar\omega_k (t'-t)}e^{ikz\cos\theta },\\
\end{split}
\end{equation}
where spherical coordinates $(k,\theta, \varphi)$ have been adopted with the zenith direction oriented along the  $\hat{z}$ axis (recall that $(u_k+v_k)^2=E_k/(\hbar\omega_k)$). 

We can then  compute  the energy level shift of the impurity, $\delta E(z)$, at  second order in $g_-$, finding (see for instance Ref. [35] of the main text)
\begin{equation}\label{intCasBog}
\begin{split}
\delta E^g(z)&=\lim_{(t\to\infty,~t_0\to-\infty)}\frac{-ig^2_-}{2}\int^t_{t_0}dt'\left(\langle \delta\rho(\mathbf{r},t)\delta\rho(\mathbf{r}',t')\rangle\langle g|\sigma_x(t)\sigma_x(t')|g\rangle-\langle \delta\rho(\mathbf{r}',t')\delta\rho(\mathbf{r},t)\rangle\langle g|\sigma_x(t')\sigma_x(t)|g\rangle\right)=\\
&=\frac{\rho_0g_-^{2}}{2}\frac{1}{(2\pi)^{2}}\int_{0}^{\infty}dk\frac{k^{2}}{\hbar^{3}}\frac{E_{k}}{\omega_k}\frac{\sin(2kz)}{kz}\frac{1}{\omega_{k}+\omega_{0}}=\\
&=_{(k=\frac{1}{\xi}y)}\frac{g_-^{2}}{4}\frac{\rho_0}{(2\pi)^{2}}\frac{1}{\hbar\mu}\frac{1}{\xi^2z}\int_{0}^{\infty}dy\frac{y^{2}}{\sqrt{1+y^{2}}}\frac{1}{y\sqrt{1+y^{2}}+\frac{\hbar\omega_{0}}{2\mu}}\sin\left(\frac{2zy}{\xi}\right),
\end{split}\end{equation}
where we integrated over the angular variables ($\theta, \varphi$). 
This is the position-dependent contribution to the Lamb Shift of a ground-state state atom in the presence of a reflecting plate, which is also commonly referred as atom-plate Casimir interaction.
In Eq. \eqref{intCasBog} we neglected a distance-independent energy shift which does not contribute to the Casimir force.

Setting $\omega_0=0$ in Eq. \eqref{intCasBog}, we find $\delta E(z)\sim e^{-2z/\xi}$, which mirrors Eq. (57) derived in Ref. [12] for one-dimensional weakly interacting bosons. 
On the other hand, for impurities with an internal optical structure $\omega_0\neq0$, we find, introducing the integration variable $Y=\frac{2\mu y}{\hbar\omega_{0}}$, for $z\gg\xi$, 
\begin{equation}\label{shiftg}
\delta E^g(z)\simeq \frac{\rho_0}{(2\pi)^2}\frac{g_-^2}{16}\frac{(\hbar\omega_0)^2}{\hbar\mu}\frac{1}{z}\frac{1}{(\xi\mu)^2}\int_0^\infty dY\frac{Y^2}{Y+1}\sin(\zeta Y)\equiv \frac{A}{z} I(\zeta),
\end{equation}
where $A=\frac{\rho_0}{(2\pi)^2}\frac{g_-^2}{16}\frac{\hbar\omega^2_0}{\mu}\frac{1}{(\xi\mu)^2}$ and $\zeta\equiv\frac{z}{\xi}\frac{\hbar\omega_0}{\mu}=\frac{z\omega_0}{c_s}$. 
We  then compute $I(\zeta)\equiv-\frac{\partial^2}{\partial\zeta^2}\mathcal{I}(\zeta)$, with $\mathcal{I}(\zeta)=\int_0^\infty dY\frac{1}{Y+1}\sin(\zeta Y)$, and we find the two asymptotic expressions ${I}(\zeta)\simeq_{\zeta\ll1}1/\zeta +O(\zeta^0)$ and $I(\zeta)\simeq_{\zeta\gg1}1/\zeta^3+O(1/\zeta^5)$, which in turn dictates the following asymptotic behaviour for the Lamb Shift (or Casimir potential)

\begin{equation}\label{casimint}
\delta E(z)=\begin{cases}
-\frac{\rho_0g_-^2}{(2\pi)^2}\frac{\omega_0\xi}{\hbar^2c_s^2}\frac{1}{z^2}, & \xi\ll z\ll c_s/\omega_0,\\
-\frac{2\rho_0g_-^2}{(2\pi)^2}\frac{\xi}{\omega_0\hbar^2}\frac{1}{z^4}, & z\gg c_s/\omega_0,
\end{cases}
\end{equation}
where we assumed $\hbar\omega_0\ll\mu$ (or $\Omega\ll1$).

From the expressions \eqref{casimint}, the force $F^g_z(z)=-\frac{\partial( \delta E^g (z))}{\partial z}$ in near and far zone of the main text follows.  

\subsection{Comparison with scalar EM Casimir force}
In the scalar EM case, the interaction term  $V=\lambda\phi(\mathbf{r})\sigma^x$, is written in terms of the scalar field $\phi(\mathbf{r})$, whose correlation function (the analogue of Eq. \eqref{correlatore}) in three dimensions  reads
\begin{equation}\label{correlatoreEM}
\begin{split}
\langle \phi(\mathbf{r},t)\phi(\mathbf{r}',t'))\rangle&=
-\int_0^{2\pi} d\varphi\int_0^{\pi} d\theta\sin\theta\int_0^\infty \frac{dk}{(2\pi)^3} k^2\frac{1}{\Omega_k} e^{i\hbar\Omega_k (t'-t)}e^{ikz\cos\theta },\\
\end{split}
\end{equation}
where $\Omega_k=ck$ ($c$ is the speed of light).
 The $z$-dependent component of the energy level shift of a ground state atom is computed within second order perturbation theory [35]  in $\lambda$ (using the atomic correlation function \eqref{atomcorr})
\begin{equation}
\delta E^g(z)\propto\frac{\lambda^2}{z}\int_0^\infty dk \frac{\sin{2kz}}{\Omega_k+\omega_0}.
\end{equation}

In near zone ($z\ll\omega_0/c$), we then find $\delta E^g(z)\sim 1/z$ ($F^g_z(z)\sim 1/z^2$), while in far zone $z\gg\omega_0/c$, the Casimir interaction potential becomes $\delta E^g(z)\sim 1/z^2$ ($F^g_z(z)\sim 1/z^3$).

Therefore, in near zone, the force felt by an impurity embedded  in a BEC is stronger than its scalar EM counterpart, while in far zone the situation is reversed.
The different scaling behaviour is traced back to the spectral weight in  the Bogolyubov  ($E_k/(\hbar\omega_k)\sim_{(k\xi\ll1)}k$) and in the scalar EM case ($1/\Omega_k\sim 1/k$), intervening respectively in Eqs.  \eqref{correlatore} and \eqref{correlatoreEM}.

\subsection{Other scaling behaviours of the Casimir force}
For an atom at rest in its excited state, $|e\rangle$, we find again, for $z\gg\xi$,
\begin{equation}
\delta E^{e}(z)\simeq\frac{\rho_0}{(2\pi)^2}\frac{g_-^2}{16}\frac{(\hbar\omega_0)^2}{\hbar\mu}\frac{1}{z}\frac{1}{(\xi\mu)^2}\mathcal{E}(\zeta),
\end{equation} 
with $\mathcal{E}(\zeta)\equiv \int_0^\infty dY\frac{Y^2}{Y-1}\sin(\zeta Y)$.
We  continue $\mathcal{E}(\zeta)$ to  complex values, and we choose an integration path forming a quarter of circle in the first quadrant of the complex plane. 
In contrast to Eq. \eqref{shiftg}, we now have a resonance pole at $Y=1$ (or in dimensionful units at $k=\omega_0/c_s$), accounting for the emission of a Bogolyubov quantum with energy equal to the  level spacing between the ground state and the excited state of the impurity. 
For $\zeta\gg1$ we find the leading behaviour $\mathcal{E}(\zeta)\simeq \pi \cos(\zeta)+...$ (a standard signature of the resonance pole, confront again with Ref. [35]), and therefore, after taking the spatial derivative $F^g_z(z)=-\frac{\partial( \delta E (z))}{\partial z}$, the expression of the force for a static excited impurity reported in the main text.\\

We now briefly outline the computation of the supersonic Casimir force.
{In the reference frame comoving with the uniformly moving impurity at speed $v$, we need to take into account that the Bogolyubov mode frequencies are Doppler shifted as $\omega_k-\mathbf{k}\cdot\mathbf{v}$. This leads to the following expression for the $z$-dependent component of the density two-point function, evaluated on the vacuum state of the field}
\begin{equation}\label{correlatoress}
\langle \delta\rho(\mathbf{r},t)\delta\rho(\mathbf{r}',t'))\rangle=
-\rho_0\int_0^{2\pi} d\varphi\int_0^{\pi} d\theta\sin\theta\int_0^\infty \frac{dk}{(2\pi)^3} k^2 (u_k+v_k)^2e^{(i\hbar\omega_k -i\hbar kv\cos\varphi\sin\theta)(t'-t)}e^{ikz\cos\theta }.
\end{equation}
Eq. \eqref{correlatoress} is written assuming the impurity moves parallel to the plate at a distance $z$ from it, with velocity $v$, therefore in the right hand side only a  $z$ dependence occurs.
Working out again the energy level shift of the ground state at second order in $g_-$, we find
\begin{equation}\label{ssshift}
\delta E^g(z)=\frac{\rho_0g_-^2}{(2\pi)^3}\frac{1}{\hbar^2\xi}\int_0^\infty dx\int_0^1 da\int_0^{2\pi}d\varphi\frac{x^3}{\xi^2\sqrt{x^2+1}}\frac{1}{\frac{2\mu}{\hbar}x\sqrt{x^2+1}+\omega_0-x\frac{v\cos{\varphi}}{\xi}\sqrt{1-a^2}}\cos\Big(ax\frac{z}{\xi}\Big),
\end{equation}
where $x$ is a  dimensionless momentum integration variable  and $a\equiv\cos\theta$; 
in the  limit $z\gg\xi$, this expressions yields 
\begin{equation}\label{integrale}
\delta E^g(z)=\frac{\rho_0g_-^2}{(2\pi)^3}\frac{1}{\hbar^2\xi^2}\frac{1}{z}\int_0^\infty dx\int_0^{2\pi}d\varphi\frac{x^2}{\frac{2\mu}{\hbar}x+\omega_0-\frac{x}{\xi}v\cos\varphi}\sin\Big(\frac{xz}{\xi}\Big).
\end{equation}
We now recall  that
\begin{equation}\label{lemma}\int_0^{2\varphi}d\varphi\frac{1}{A-B\cos\varphi}=\begin{cases}
\frac{2\pi}{\sqrt{A^2-B^2}}, & A>B,\\
0 & A<B,
\end{cases}
\end{equation} 
where  the integration has been extended to the complex plane via $z=e^{i\varphi}$: when $A<B$, both poles of the denominator lie inside the  unit circle and their residues cancel each other, while, for $A>B$, just only one of them lie inside, giving the result in the first line of \eqref{lemma}.
Applying this lemma to the $\varphi$-integration in \eqref{integrale}, we find a non-vanishing result only when $\frac{2\mu}{\hbar}x+\omega_0>\frac{x}{\xi}v$, or equivalently $\frac{1}{\xi}(v-c_s)<\omega_0$.
This implies that for $c_s>v$ the integral domain in Eq. \eqref{integrale} is $0<x<\infty$, while for $v>c_s$, it is $0<x<\xi\omega_0/(v-c_s)$. %
The presence of an upper integration bound is at origin of the different scaling in the supersonic regime reported in the main text, which follows after lengthy algebra from Eq. \eqref{integrale}.

\end{document}